\documentclass{article}

\usepackage{amsmath,amssymb,amsthm,multirow,bbm,cite,threeparttable,booktabs}
\usepackage{spconf,graphicx,subfigure,comment,balance,caption,indentfirst}
\usepackage{xcolor}
\newcommand{\bh}{\boldsymbol{h}}
\newcommand{\bz}{\boldsymbol{z}}

\newcommand{\bd}{\boldsymbol{d}}

\newcommand{\bx}{\boldsymbol{x}}
\newcommand{\bX}{\boldsymbol{X}}
\newcommand{\bg}{\boldsymbol{g}}
\newcommand{\bA}{\boldsymbol{A}}

\newcommand{\be}{\boldsymbol{e}}
\newcommand{\bp}{\boldsymbol{p}}
\newcommand{\bH}{\boldsymbol{H}}

\newcommand{\mH}{\mathrm{H}}
\newcommand{\glob}{\text{glob}}
\usepackage{color}
\def\cred{\textcolor{blue}}  
\def\cblue{\textcolor{black}}  

\newcommand{\bpsi}{\boldsymbol{\psi}}

\title{Distributed pressure matching for Personal Sound Zone Control \\ using diffusion adaptation}

\name{Mengfei Zhang$^{*\dagger}$, Junqing Zhang$^*$, Jie Chen$^*$, C\'edric Richard$^\dagger$\thanks{The work of M. Zhang was supported partly by the China Scholarship Council and partly by the Innovation Foundation for Doctor Dissertation of Northwestern Polytechnical University. The work of J. Chen is partially supported by NSFC grant 61671382, Shaanxi Key Industrial Innovation Chain Project 2022ZDLGY01-02, and Xi’an Technology Industrialization Plan  XA2020-RGZNTJ-0076. The work of C. Richard was funded in part by the PIA program under its IDEX UCAJEDI project (ANR-15-IDEX-0001) and by ANR under grant ANR-19-CE48-0002.}}
\address{\small $^*$ 
\small School of Marine Science and Technology, Northwestern Polytechnical University, China \\
		\small  $^\dagger$ Universit\'e  C\^ote d'Azur, CNRS, OCA, France}

\begin{document}
\ninept
\maketitle
%
\begin{abstract}
Personal sound zone (PSZ) systems, which aim to create listening (bright) and silent (dark) zones in neighboring regions of space, are often based on time-varying acoustics. Conventional adaptive-based methods for handling PSZ tasks suffer from the collection and processing of acoustic transfer functions~(ATFs) between all the matching microphones and all the loudspeakers in a centralized manner, resulting in high calculation complexity and costly accuracy requirements. This paper presents a distributed pressure-matching (PM) method relying on diffusion adaptation (DPM-D) to spread the computational load amongst nodes in order to overcome these issues. The global PM problem is defined as a sum of local costs, and the diffusion adaption approach is then used to create a distributed solution that just needs local information exchanges. Simulations over multi-frequency bins and a computational complexity analysis are conducted to evaluate the properties of the algorithm and to compare it with centralized counterparts.
\end{abstract}
%
\section{Introduction}
\label{sec:intro}
\noindent Personal sound zone (PSZ) aims to generate individual zones for users within a spatial control region by employing a loudspeaker array~\cite{betlehem2015personal}. This can
be used in a wide range of commercial audio applications, including neckband headset~\cite{jeon2020personal}, car cabin audio~\cite{so2019subband}, mobile devices~\cite{cheer2013practical}, television sound systems~\cite{galvez2014personal}, to cite a few.
Among several methods adopted for sound zone generation, there are two typical approaches, namely the acoustic contrast control~(ACC)~\cite{choi_02,shin_10,coleman_14j2} methods and pressure matching (PM)~\cite{olivieri2017generation,moles2019providing,moles2022weighted,zhang2022robust} methods. The ACC algorithms maximize the energy ratio between the bright and dark zones. The PM algorithms, on the other hand, try to minimize the mean-square error (MSE) of the reproduced sound field in the zones compared to the target sound field. Note that these methods 
should be implemented with online estimates of room impulse responses~(RIR), which are often time-variant~\cite{spors2007active,brannmark2013compensation}. 

Existing adaptive PSZ algorithms employ centralized approaches to gather and process the acoustic transfer functions (ATFs) between all loudspeakers and all {matching} points~\cite{vindrola2021use}. However, the practicality of these centralized solutions is limited in large-scale applications and distributed system deployments due to their high computational complexity. To address this, network-wide ACC-based methods have been proposed, which leverage a wireless acoustic sensor and actuator network (WASAN) to distribute the computational burden across nodes~\cite{pinero2017feasibility,van2021distributed}. In the method described in~\cite{pinero2017feasibility}, each zone is treated as being covered by a single acoustic node, composed of $L$ loudspeakers and one microphone. However, this approach requires significant processing capacity for each node. Alternatively, in \cite{van2021distributed}, a distributed adaptive ACC algorithm using a gradient-based generalized eigenvalue decomposition (GEVD) approach is devised to solve the centralized problem, achieving comparable performance to its centralized counterpart. Nevertheless, this method relies on a root node to compute the global gradient vector and disseminate it to all other nodes through a communication tree. If a communication or computation failure occurs on the root node, it can result in the failure of the entire network. 
To overcome these limitations, the diffusion adaptation strategy~\cite{Lopes2008Diffusion,Sayed2014Diffusion,Chen2015diffusion} presents an attractive solution with enhanced adaptation performance and wider stability ranges. It operates without relying on a root node and allows each node to share its local estimates with immediate neighbors, optimizing a global cost collaboratively. Moreover, the optimization objective of the PM method, which is a quadratic form, can be decomposed into a sum of local cost functions, making it more suitable for exploiting the benefits of diffusion strategies. These factors motivated us to develop a distributed PM algorithm using diffusion adaptation.

In this paper, we examine a real-world PSZ application scenario where the ATFs exhibit time-varying characteristics and are subject to regular perturbations caused by physical and environmental factors~\cite{zhu2016robust,zhang2022robust}. Within this context, we propose a distributed PM algorithm designed for networks. Our approach considers a system architecture where each node consists of multiple microphones, multiple loudspeakers, and a local processor with communication and computing capabilities.
Moreover, we impose the constraint that each node $k$ can only access its local data, specifically the measured ATFs between its microphones and all the loudspeakers. Subsequently, we decompose the global cost function of the PM algorithm across the nodes to address the optimization problem in a distributed manner. We conduct simulations to validate the proposed distributed pressure matching via diffusion adaptation (DPM-D) algorithm across multiple frequency bins.
This approach stands out significantly from the existing literature, which predominantly emphasizes the use of acoustic contrast control (ACC)-based distributed methods. Our proposed method surpasses the limitations of the distributed ACC approach, which necessitates a root node for communication and computation.

\noindent\textbf{Notation.} Normal font letters $x$, boldface small letters $\bx$ and capital letters $\bX$ denote scalars, column vectors and matrices, respectively. $\mathbb{C}$ denotes the complex field. The transpose and Hermitian transpose of a vector or matrix are denoted by $(\cdot)^{\top}$ and $(\cdot)^\mH$, respectively. $(\cdot)^\ast$ denotes the complex conjugation operator. $\mathbbm{1}_N$ denotes the all-one vector of size $N$. $\mathcal{N}_{k}$ denotes the set of one-hop neighboring nodes of node $k$, including $k$, \cblue{and $|\mathcal{N}_{k}|$ denotes the cardinality of $\mathcal{N}_{k}$. }

\begin{figure*}[htb]
	\centering
		\subfigure[] {\label{fig:sm}{\includegraphics[width=6cm]{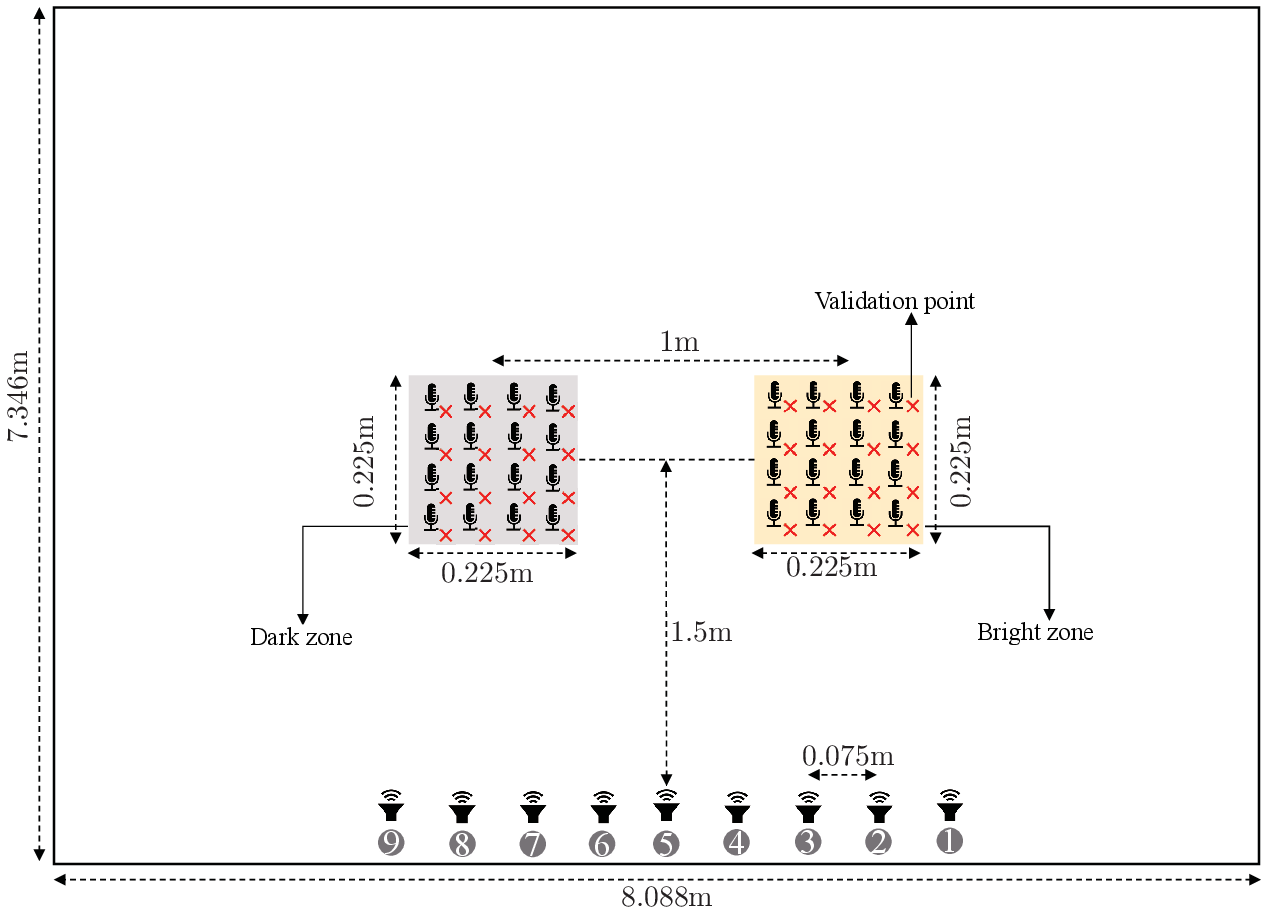}}}\hspace{0.5cm}
		\subfigure[] {\label{fig:s1}{\includegraphics[width=4.9cm]{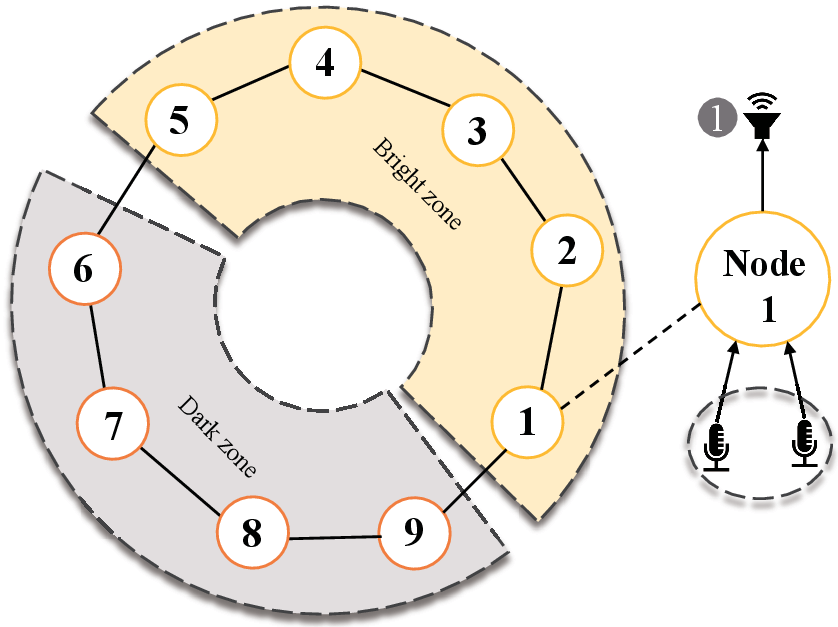}}}\hspace{.5cm}
		\subfigure[] {\label{fig:s2}{\includegraphics[width=5.68cm]{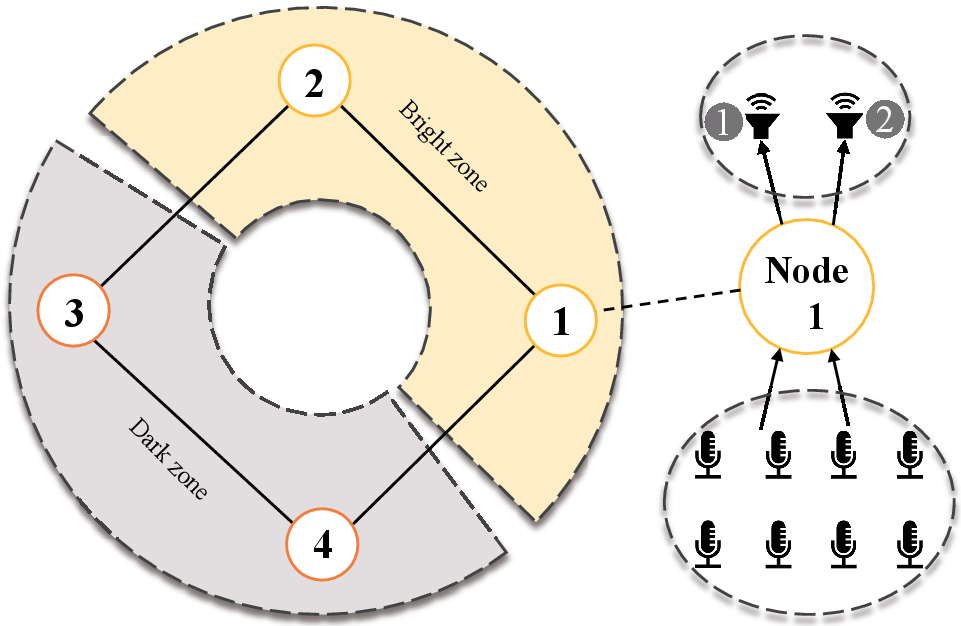}}}
	\vspace{-4mm}\caption{Simulated PSZ system and examples of network topologies for distributed systems. (a) \cblue{Plan view of the system geometry within a room, where $\color{red}{\times}$ represents the position of each validation point. The PSZ system is composed of $9$ loudspeakers, and $16$ microphones in the bright and in the dark zone.} (b) System 1: a network of $9$ nodes, where each node possesses $2$ or $4$ microphones and $1$ speaker; specifically, each node $l$ controls its own speaker $l$ and possesses $2$ or $4$ microphones in the bright or in the dark zone depending if it is assigned to the bright or to the dark zone.
  (c) System 2: same as System 1 except that the network now has $4$ nodes, and each node possesses $8$ microphones in the bright or in the dark zone, and controls $2$ or $3$ speakers among $9$ available.}
	\label{fig:sys_mod}\vspace{-6mm}
\end{figure*}
\section{Problem formulation and Centralized adaptation strategy }
\label{sec:format}

\subsection{System model}
As shown in~Fig.~\ref{fig:sm}, we assume  
an array of $L$ loudspeakers 
and two spatial regions with $M$ control points in total, delimited by $M_{\mathrm{b}}$ (bright zone) and $M_{\mathrm{d}}$ (dark zone) control points, respectively. The objective of this system is to render a target sound field in the bright zone with {minimal interference} in the dark zone. The subscripts ${}_{\mathrm{b}}$ and ${}_{\mathrm{d}}$ denote the bright and dark zones, respectively. 

With the frequency domain approach, the sound pressure $p_{m}$ at the $m$-th control point is given by:
  \setlength{\abovedisplayskip}{2pt}
\setlength{\belowdisplayskip}{2pt}
\begin{align}
\label{eq:def_p}
p_{m}(f) = \sum_{l =1}^{L}h_{m,l}(f)g_l(f) = 
\bh_{m}(f)\bg(f),
\end{align}
where $f$ is the frequency, $h_{m,l}(f) \in 
\mathbb{C}$ denotes the ATF between the $l$-th loudspeaker to the $m$-th control point, $g_l(f) \in \mathbb{C}$ denotes the loudspeaker control filter, and: 
\begin{align}
\label{eq:def_h&w}
\bh_{m}(f) &= [h_{m,1}(f)\,\cdots\, h_{m,L}(f)], \\
\bg(f) &= [g_1(f)\, \cdots\, g_L(f)]^\top.
\end{align}
We combine the ATF matrices of the bright and dark zones into a matrix $\bH(f)$ as follows:
\begin{align}
\label{eq:def_ATF}
\bH(f) = \begin{bmatrix}\bH_{\mathrm{b}} \\ \bH_{\mathrm{d}}\end{bmatrix}= \begin{bmatrix}
\bh_{1}(f)\\  \vdots \\ \bh_{M_{\mathrm{b}}}(f)\\ \bh_{M_{\mathrm{b}}+1}(f)\\ \vdots \\ \bh_{M}(f)
\end{bmatrix},
\end{align}
where  the first $M_{\mathrm{b}}$ row vectors of $\bH$ denotes the ATFs for the bright zone. $\bH_{\mathrm{d}}$, consisting of  the other $M_{\mathrm{d}} = M - M_{\mathrm{b}}$ row vectors of $\bH$, denotes the ATFs for the dark zone. 
Likewise, the vector $\bp(f)$ containing the sound pressure at the $M$ control points is defined as:
\begin{align}
\label{eq:def_V_p}
\bp(f) &= [p_{1}(f) \cdots p_{M_{\mathrm{b}}}(f),  p_{M_{\mathrm{b}}+1}(f)\cdots p_{M}(f)]^\top \notag \\
&= \bH(f) \bg(f),
\end{align}
and the desired signal $\bd(f)$ at all the $M$ control points as: 
\begin{align}
\label{eq:def_V_d}
\bd(f) = [d_{1}(f) \cdots d_{M_{\mathrm{b}}}(f),d_{M_{\mathrm{b}}+1}(f)\cdots d_{M}(f)]^\top.
\end{align}
From \eqref{eq:def_V_p} and \eqref{eq:def_V_d}, the estimated error at control points is given by:
\begin{align}
\label{eq:def_error}
\be(f) = \bp(f) - \bd(f).
\end{align}
\subsection{Centralized adaptive PM}
ATFs are usually measured beforehand
, that is, during a pre-calibration stage. Then they are used for the control filter calculation. Nonetheless, perturbations are unavoidable during the actual measurement of ATF because of, e.g., a position mismatch of sensors, RIR variation caused by changes in room temperature and humidity, changes in the electroacoustic response, and background noise in the ATF measurement procedure, etc. It is reasonable to continuously estimate the control filter $\bg$ with the ATFs being updated over time. We rewrite the estimated error of \eqref{eq:def_error} at the $n$-th time block as: 
\begin{align}
\label{eq:error}
\be(n,f) = \bp(n,f) - \bd(n,f).
\end{align}
We consider the mean-square error criterion $J(n,f) \triangleq \|\be(n,f)  \|^2$ to formulate the problem of frequency-domain adaptive PM~\cite{vindrola2021use}:
\begin{equation}
    \label{eq:cost}
    \bg^o(n,f)=\arg\min_{\bg}J(n,f).
\end{equation}
Taking the derivative
of $J(n,f) $ w.r.t. $\bg$ and considering the complex least mean squares (LMS) algorithm \cite{widrow1975complex}, the adaptive update equation can be written as:
\begin{align}
\label{eq:update_w}
\bg(n+1,f) =& \bg(n,f) - \mu\bH^\mH(n,f)\be(n,f)  ,
\end{align}
where $\mu>0$  is the step size. Inspecting \eqref{eq:update_w}, we observe that the error signals as well as the ATFs between each control point and all the loudspeakers are necessary for computing the control filter at each iteration. This means that \eqref{eq:update_w} must be processed in a centralized manner\cite{sayed2014adaptation}.

\section{Distributed pressure matching via  diffusion adaptation}
\label{sec:dPM}

\subsection{Distributed diffusion adaptation strategy}

Consider solving \eqref{eq:cost} in a collaborative and distributed manner. Before devising the proposed algorithm, we briefly introduce the diffusion adaptation strategy. Consider a connected network composed of $N$ nodes. The problem is to estimate an unknown complex vector $\bg^o \in\mathbb{C}^{L\times 1}$ such that the following global cost is minimized:
\begin{align}
\label{eq:glob_cost}
J^\glob(\bg) = \sum_{k=1}^{N}J_k(\bg),
\end{align}
where $J_k(\bg)$ denotes a real-valued function accessible to node $k$ that is considered to be convex.

 The typical adapt-then-combine (ATC) diffusion LMS strategy is written in the following form \cite{Sayed2014Diffusion}:
 \begin{align}
\label{eq:A-diffusion}
\bpsi_{k,n+1} &= \bg_{k,n} - \mu[\hat{\nabla} J_{k}(\bg_{k,n})]^\ast,\\
\bg_{k,n} &= \sum_{\ell \in \mathcal{N}_k}a_{\ell k}\bpsi_{\ell,n+1},
\label{eq:C-diffusion}
\end{align}
where $\hat{\nabla}J_{k}(\bg_{k,n})$ denotes a stochastic approximation for the true local gradient $\nabla J_k(\bg_{k,n})$, the nonnegative coefficients $a_{\ell k}$ denote the $(\ell,k)$-th entries of a left-stochastic matrix $\bA$, satisfying:
	\begin{align}
	&\bA^\top\mathbbm{1}_N =\mathbbm{1}_N,a_{\ell k}=0,\;\;\; {\rm{if}}\;\; \ell\notin {\cal{N}}_k \label{eq:Constraints}
	\end{align}
to ensure convergence in the mean sense towards $\bg^o$,
and $\mu_k$ is a positive step size at node $k$.
\subsection{Distributed PM via diffusion LMS}
To facilitate the presentation of the proposed strategy, we define the network model as follows. We focus on a $N$-node distributed personal sound zone network. Node $k$ is a module consisting of one or more microphones in the control zone, one or more loudspeakers, and a processor with communication and computation capabilities. A network of multi-channel nodes capable of processing a variety of microphones' and loudspeakers' data is depicted in Fig.\ref{fig:s1}-\ref{fig:s2}. Note that, in practice, the system architecture matches the characteristics of the application at hand (number of processors, number of microphones and loudspeakers, room size, etc.).

For simplicity, in this paper, we consider the case where each node is equipped with $M_k$ microphones and $L_k$ loudspeakers.  We write $M=\sum_{k=1}^N M_k$ and $L=\sum_{k=1}^N L_k$. Let $\{\mathcal{C}_k\}_{k=1}^{N}$ be a partition of the set of indices $\mathcal{C} = \{1,\cdots, M\}$, specifically,
\begin{align}
\label{eq:Ck}
\bigcup_{k=1}^N\mathcal{C}_k = \mathcal{C}, \;\;\;\mathcal{C}_k \cap \mathcal{C}_\ell = \emptyset,\; \text{if} \;k\neq \ell,
\end{align}
where $\mathcal{C}_k$ denotes the set of indices of microphones at node $k$. We further rewrite the global cost function \eqref{eq:cost} as \eqref{eq:glob_cost}:
\begin{align}
\label{eq:dis_cost}
\bg^o=\arg \min_{\bg} J^\glob(\bg) = &\sum_{k=1}^NJ_k(\bg)
\end{align}
where we omitted the time block $n$ and the frequency $f$ for simplicity. 
The local cost function $J_k(\bg)$ at node $k$ is written as:
$$J_k(\bg) = \sum_{\ell \in\mathcal{C}_k} e_\ell^2$$
where  the $\ell$-th entry of vector $\be$, i.e., the error signal at the $\ell$-th microphone, is given by $e_\ell = p_\ell - \bH(\ell,:)\bg$,
with $\bH(\ell,:) \in \mathbb{C}^{L\times 1}$ the $\ell$-th row of $\bH$, which denotes the ATFs from the $\ell$-th microphone to all $L$ loudspeakers at node $k$. 
We force each node $k$ to have access only to its local measurements, which means that node $k$ estimates a local version $\bg_k$ of the control filter $\bg$ without exchanging the row data of $\bH$ with other nodes.
Taking the derivative of $J_k(\bg)$ w.r.t. $\bg$:
\begin{align}
\label{eq:gra_Jk}
\nabla J_k(\bg)= \sum_{\ell \in \mathcal{C}_k} 2\bH^\mH(\ell,:)e_\ell. \end{align}
Considering the diffusion LMS strategy in \eqref{eq:A-diffusion}-\eqref{eq:C-diffusion}, we can
address problem~\eqref{eq:dis_cost} in a distributed manner using only local measurements available at each node $k$: 
\begin{align}
\label{eq:A_diffusion_szc}
\bpsi_{k,n+1} &= \bg_{k,n} - \mu_k \sum_{\ell \in \mathcal{C}_k} \bH^\mH(\ell,:)e_\ell,\\
\bg_{k,n+1} &= \sum_{\ell \in \mathcal{N}_k}a_{\ell k}\bpsi_{\ell,n+1},
\label{eq:C_diffusion_szc}
\end{align}
where  coefficients $\{a_{\ell k}\}$ satisfy the conditions in \eqref{eq:Constraints}. The local
estimate $\bg_{k,n}$ of the unknown control filter $\bg^o$ is defined by
\begin{align}
\label{eq:loc_gk}
\bg_{k,n} = [\bar{\bg}_{1,k,n}^\top\cdots\bar{\bg}_{\ell,k,n}^\top,\cdots,\bar{\bg}_{N,k,n}^\top,]^\top,
\end{align}
where $\bar{\bg}_{\ell,k,n}$ consists of the control weights coefficients of the $L_\ell$ loudspeakers at node $\ell$ and time index $n$, and $\bg_{k,n} $ are the control filter weights of all the $L$ loudspeakers. The solution of \eqref{eq:dis_cost} is the optimal control filter~$\bg^o = [g_1^o, \cdots, g_L^o]^\top$.

 \begin{figure*}[htb]
	\centering
	\subfigure[] {\label{fig:mse}{\includegraphics[width=5.8cm]{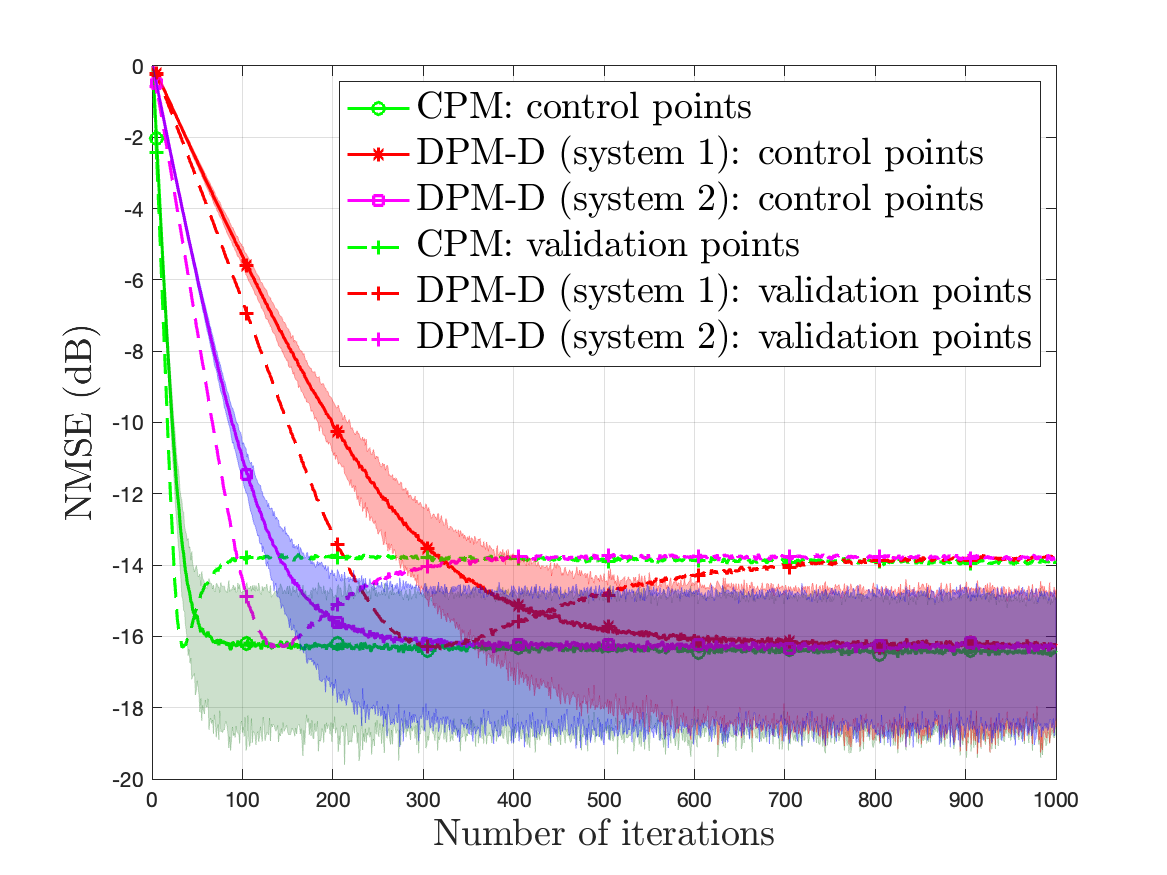}}\vspace{-4mm}} \hspace{1cm}
	\subfigure[] {\label{fig:ac}{\includegraphics[width=5.8cm]{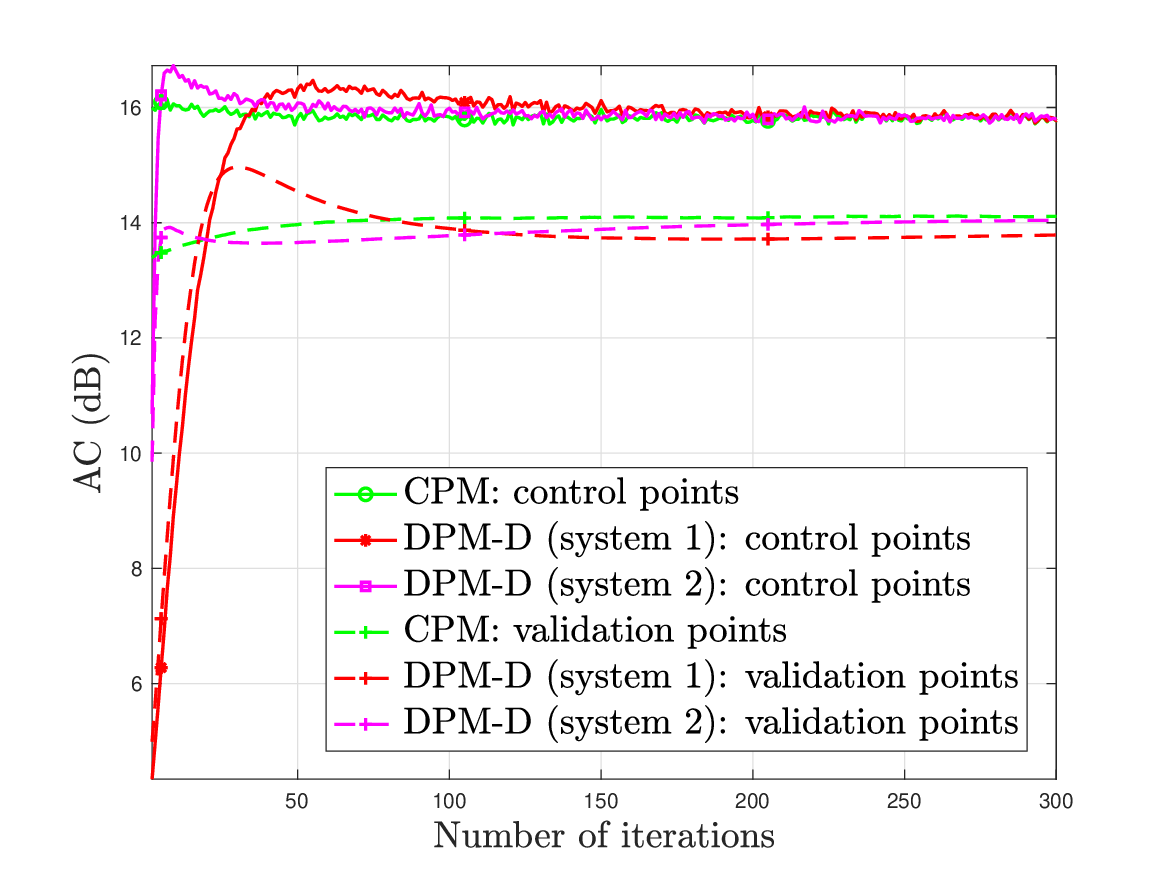}}\vspace{-4mm}}
	\vspace{-4mm}\caption{\cblue{Comparison of the NMSE and AC learning curves. Shaded regions in (a) represent the three standard deviations of the estimates.}}	\label{fig:mse_ac}\vspace{-3mm}
	\end{figure*}

 \begin{figure*}[htb]
	\centering
	\subfigure[] {\label{fig:mul_mse}{\includegraphics[width=5.8cm]{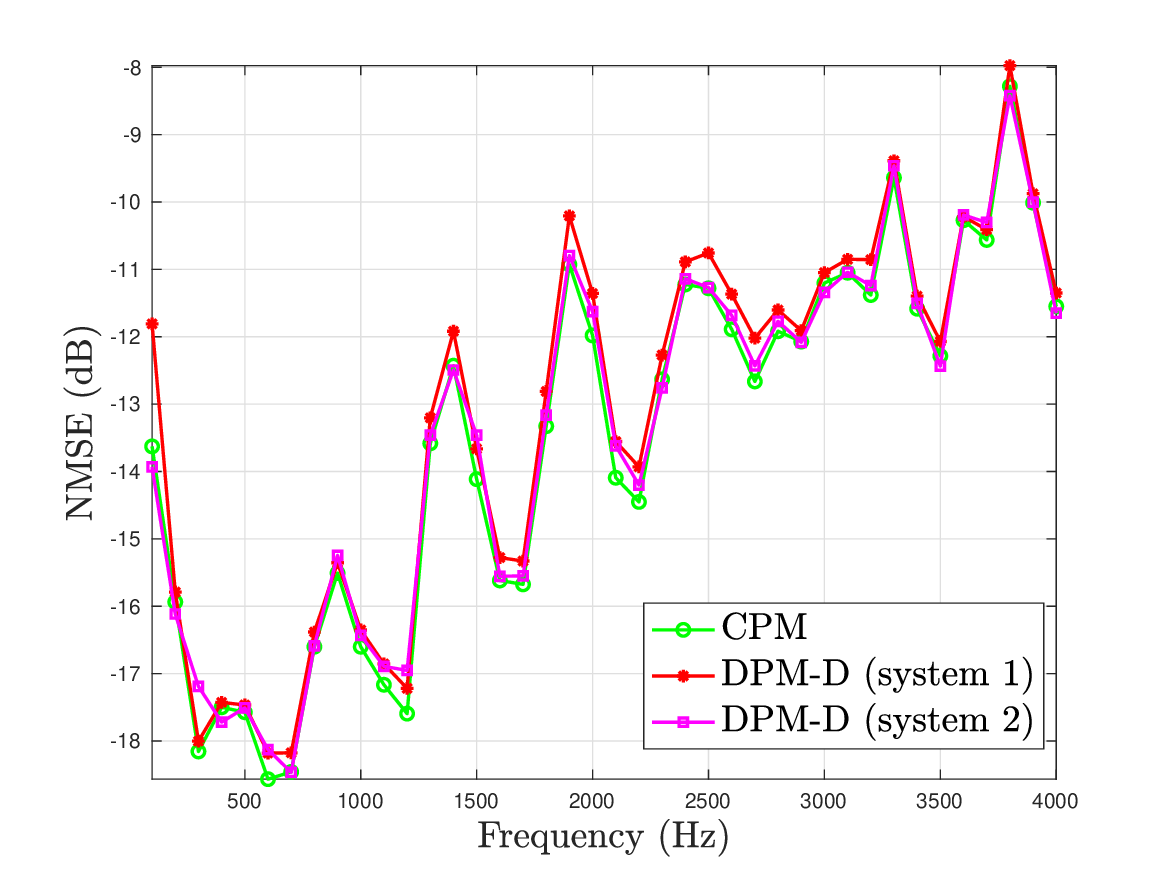}}} \hspace{1cm}
	\subfigure[] {\label{fig:mul_ac}{\includegraphics[width=5.8cm]{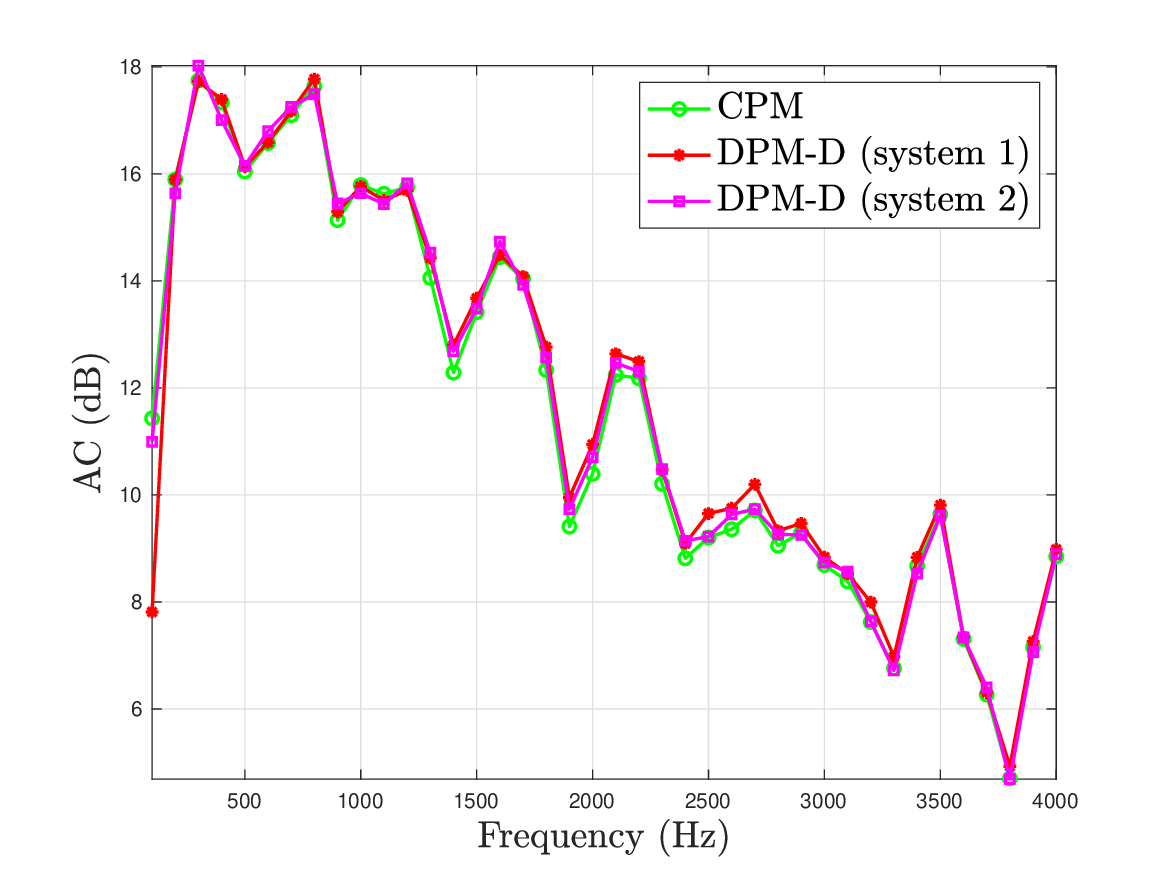}}}
	\vspace{-5mm}\caption{Multi-frequency control performances of the proposed DPM-D algorithm for Systems 1 and 2 compared with the CPM algorithm \cblue{ on control points}. (a) NMSE on frequency bins
after $5000$ iterations; (b) AC on frequency bins
 after $5000$ iterations.}	\label{fig:mul_mse_ac}\vspace{-3mm}
	\end{figure*}
 \cred{
 \begin{table*}[!t]
\centering
\caption{Comparison of computational complexity}
\setlength{\tabcolsep}{4.2mm}{
\renewcommand\arraystretch{1.2}\vspace{-3mm}
\begin{tabular}{lll}
\hline\hline
    Algorithms & Additions & Multiplications\\
\hline
     CPM& $(M+L)\times {F}\log_2 F+M\times L$ &$(M+L)\times \frac{F}{2}\log_2 F+(M+1)\times L $\\
     Proposed DPM-D & $(M_k+L_k)\times {F}\log_2 F+(|\mathcal{C}_k|+|\mathcal{N}_k|-1)\times L$&$(M_k+L_k)\times\frac{F}{2}\log_2 F+(|\mathcal{C}_k| +|\mathcal{N}_k|+1)\times L$\\
\hline
\hline
\end{tabular}}
\label{tab:ComputationalComplexity} \vspace{-3mm}
\end{table*}
}
\section{SIMULATIONS}
\label{sec:simu}
In this section, we validate the proposed Distributed PM method via diffusion LMS \eqref{eq:A_diffusion_szc}-\eqref{eq:C_diffusion_szc}, denoted by DPM-D. The algorithm is compared with its centralized counterpart denoted by CPM. All curves were obtained by averaging over 100 Monte Carlo runs.
\subsection{Simulation setup}
In the simulation, the room environment was generated with the RIR generator toolbox~\cite{habets2006room}. As depicted in Fig.~\ref{fig:sm}, a rectangular room of size $8.088 \;\text{m} \times7.346\; \text{m} \times \;2.865 \;\text{m}$ was modeled with $T_{60} \approx 200$ ms. The microphone and loudspeaker array were on the same plane of height $1.485$~m. A uniform linear array of $L = 9$ loudspeakers with an inter-element spacing of $0.06$~m, and two square regions of size $0.225 \;\text{m} \times0.225 \;\text{m}$ with $1$~m separation was used. In both the bright and dark zones, a grid of $4 \times 4$ microphones with an inter-element distance of $0.075$~m  was used to capture the sound pressure. 
\cblue{A validation point was positioned $0.707$ cm from each control microphone.}

A sampling rate of $8$ kHz and a window length of $3200$ samples were employed. In  order to simulate the perturbations during the ATF measurement process, the ATFs were perturbed by a zero-mean white Gaussian noise with variance $0.0707$ at each iteration. 
In order to account for random thermodynamic fluctuations in the electrical signals generated by the microphone's components, an additive Gaussian noise $\bz(n)$ with a signal-to-noise ratio (SNR) of $20$ dB was added as $\bd(n) = \bH(n)\bg^o + \bz(n) $, where  we set the unknown variable vector $\bg^o$ to be a fixed set of variables sampled from $\mathcal{C}\mathcal{N}(0,1)$. We selected a fixed step size $\mu = \mu_k = 2.5$ for both the CPM and DPM-D algorithms for comparison. We simulated two systems with different numbers of nodes with ring topology, and specific settings of microphones and loudspeakers, as shown in Figs.~\ref{fig:s1}-\ref{fig:s2}. 

To evaluate the performance of the proposed algorithm, we considered (\romannumeral1) the Normalized Mean Square Error (NMSE) between the reproduced and the target sound fields within the bright zone: 
\begin{align}
 \label{eq:nmse}
 \text{NMSE} &=10\log_{10}\Big(\frac{\sum_{m\in M_{\mathrm{b}}}| d_m - p_m|^2}{\sum_{m\in M_{\mathrm{b}}}| d_m|^2}\Big),
 \end{align}
and (\romannumeral2) the Acoustic Contrast (AC), which represents the energy ratio  between the bright and dark zones after being controlled: 
 \begin{align}
 \label{eq:def_AC}
 \text{AC} =10\log_{10}\Big(\frac{M_{\mathrm{d}}\| \bH_{\mathrm{b}}\bg\|^2}{M_{\mathrm{b}}\|\bH_{\mathrm{d}}\bg \|^2}\Big),
 \end{align}
where we set $M_{\mathrm{b}} = M_{\mathrm{d}} = 16$ in  our simulations. 
\subsection{Simulation results}
We first considered that the desired signal for the bright zone is driven by a plane wave with frequency of $1$ kHz, while the desired signal for the dark zone is null \cite{moles2020personal}. \cblue{Fig.~\ref{fig:mse} illustrates the NMSE convergence behavior at the control and validation points for the bright zone, and Fig.~\ref{fig:ac} reports the AC at the control and validation points (but includes information from the bright and the dark zones. 
At the control and validation points, the DPM-D algorithm reaches a steady-state comparable to that of the CPM algorithm. At steady-state on control points, the NMSE and the AC are approximately equal to $-16$ dB and $16$ dB, respectively. At steady-state on validation points, the NMSE and the AC are approximately equal to $-14$ dB and $14$ dB, respectively.}
Regarding the DPM-D method, the distributed PM with System 2 achieves superior convergence performance. The reason is that nodes in System~2 use more microphones measurements during the update step \eqref{eq:A_diffusion_szc}. However, the resulting computational complexity is larger because it grows with the number of microphones maintained by each node. 

In order to examine the performance of the proposed algorithm over multi-frequency bins, we then tested the desired signal for the bright zone with frequency bins ranging from $100$ to $4000$ Hz with a step of $100$ Hz. Other experimental settings were identical to those considered before with the single-frequency signal. Fig.~\ref{fig:mul_mse_ac} shows the NMSE and AC behaviors at various frequencies.   It can be observed that the DPM-D with both systems and the CPM perform almost the same over most of frequency bins after $5000$ iterations.   Fig.~\ref{fig:mul_mse} and Fig.~\ref{fig:mul_ac} indicate that adding additional microphones to a node in a distributed system does not significantly improve its NMSE steady-state performance and its AC steady-state performance, respectively.

We further conducted an analysis of the computational complexity of the frequency domain-based PM methods, as shown in Table~\ref{tab:ComputationalComplexity}. The parameter $F$ in this table represents the number of FFT operations. We evaluated the computational load at each iteration on the centralized processor for centralized algorithms and on the processor of each node $k$ for distributed algorithms. The computational complexity was divided into two components: the FFT operation and the frequency domain processing. The results presented in Table~\ref{tab:ComputationalComplexity} demonstrate that the proposed distributed method effectively distribute the computational load among to the processors of each node, thereby reducing the communication load compared to the centralized approach. This improvement in load distribution enhances the scalability of the system.



\section{Conclusion}
\label{sec:conc}

The novelty of the work presented in this paper resides in the utilization of a distributed PM approach based on diffusion LMS for handling PSZ tasks. This approach stands out significantly from the existing literature, which predominantly emphasizes the use of acoustic contrast control (ACC)-based distributed methods. Our proposed method surpasses the limitations of the distributed ACC approach, which necessitates a root node for communication and computation, by enabling each node to independently estimate and share information with its neighboring nodes.


   \vfill
 \pagebreak

  \clearpage
  \newpage
  \balance
   \bibliographystyle{IEEEbib}
  \bibliography{refs}

\end{document}